\documentclass[aps,pra,preprint,showpacs,floatfix]{revtex4}
\usepackage{bm}
\usepackage{graphicx} 
\usepackage{amsmath}
\usepackage{subfigure}
\usepackage{epsfig}
\usepackage{xcolor}

\begin{document}
\title{Multichannel quantum defect theory for ro-vibrational transitions in ultracold molecule-molecule collisions}
\author{Jisha Hazra}
\affiliation{Department of Chemistry, University of Nevada Las Vegas, Las Vegas,
  Nevada 89154}
\author{Brandon  P. Ruzic}
\affiliation{JILA, NIST, and Department of Physics, University of Colorado, Boulder, Colorado 80309, USA}
\author{N. Balakrishnan}
\affiliation{Department of Chemistry, University of Nevada Las Vegas, Las Vegas,\
  Nevada 89154}
\author{John L. Bohn}
\affiliation{JILA, NIST, and Department of Physics, University of Colorado, Boulder, Colorado 80309,\ USA}

\date{\today}

\begin{abstract}
Multichannel quantum defect theory (MQDT) has been widely applied to resonant and non-resonant scattering in a variety of atomic collision
processes. In recent years, the method has been applied to cold  collisions with considerable success, and it 
has proven to be a computationally viable alternative to full-close coupling (CC) calculations when spin, hyperfine and external field 
effects are included. In this paper, we describe 
a hybrid approach for molecule-molecule scattering that includes the simplicity of MQDT while treating the short-range interaction 
explicitly using CC calculations. This hybrid approach, demonstrated for 
H$_2$-H$_2$ collisions in full-dimensionality,  is shown to adequately reproduce cross sections for quasi-resonant rotational and vibrational 
transitions in the ultracold (1$\mu$K) and
$\sim$ 1-10 K regime spanning seven orders of magnitude. It is further shown that an energy-independent
short-range $K$-matrix evaluated in the ultracold regime (1$\mu$K) can adequately characterize cross sections in
the mK-K regime when no shape resonances are present.  The hybrid CC-MQDT 
formalism provides an alternative approach 
to full CC calculations at considerably less computational expense for cold and ultracold  molecular scattering.
\end{abstract}


\maketitle

\section{Introduction}

Molecules in a translationally cold gas  present a particular  perspective on collisions and chemistry.  One the one hand, atoms in the 
colliding molecules exchange energy on the scale of tens to thousands of Kelvin, driven by deep potential energy surfaces.  
On such surfaces occur rotational, vibrational, and chemical transformations.  On the other hand, the ability of the molecules to 
initiate this activity is strongly dependent on behavior at the $\mu$K - mK translational energy scales of the gas.  The slowly moving molecules, 
to get close enough to react, must first negotiate their way through the long-range forces acting between them.  These forces, 
negligible at room temperature, loom large in the ultracold.  The dominance of long-rage forces had led to {\it control} over chemical 
reaction, by, for example, the simple expedient of applying a modest electric field to alter the dipole moments of molecules \cite{Ospelkaus}. 
For this reason cold molecules are seen as novel tools for probing and controlling chemistry with unprecedented resolution \cite{Krems}.

This dichotomy of energy scales presents a unique point of view for theories of molecules interacting at ultracold temperatures, which must
now account for dynamics occurring over many orders of magnitude in energy.  Luckily, the energy dichotomy relates in a direct way to 
motion on disparate spatial scales.  Specifically, the full, energy-sharing dynamics of atoms in the collision complex occurs where all 
participating atoms are close together, whereas the long-range dynamics occurs between well-delineated collision partners that are far apart. 
The business of cold collision theory is to accurately account for the relatively straightforward long-range dynamics, while 
incorporating, to the extent desirable or reasonable, the short-range dynamics.  

The separation into short- and long-range physics finds its natural expression in the multichannel quantum defect theory (MQDT), whose
origins go back to understanding spectra of Rydberg atoms \cite{Seaton1,Seaton2,Fano}, but which has been successfully extended to more general 
contexts \cite{Mies1,Greene84_PRA}, including cold collisions of atoms \cite{Mies2,Burke,Raou,Gao1,Hanna09_PRA}, atoms and ions \cite{Id,Gao2},
atoms and molecules \cite{Hud1,Hud2}, and molecule-molecule reactive scattering \cite{Id1,Id2,Gao3,Wang}.  In all cases, long-range wave 
functions are carefully constructed and then matched to a wave function that is a suitable representation of the short-range physics.  
Depending on the context, the short-range physics can be successfully treated in a schematic way by (for example) positing absorbing 
boundary conditions to represent chemical reactions \cite{Id1,Id2,Wang} or, in the case of alkali atom cold collisions, by means
of simple spin-dependent phase shifts \cite{Burke,Peres14_PRA}. 

In this article we tackle head-on the complete short-range dynamics of molecule-molecule scattering for the comparatively straightforward 
case of H$_2$+H$_2$ collisions at collision energies $\leq10$ K, where comparison with numerically accurate scattering calculations can be made. 
A main finding is that the MQDT approach can be accurate and considerably more efficient numerically, provided sufficient care is taken in
constructing the long-range wave functions.  Thus short-range and long-range dynamics can be successfully welded together in this important
prototype case where energy can be exchanged between rotational and vibrational degrees of freedom of two molecules.  The calculations presented here represent a first, necessary step toward adapting MQDT methods to the broader problem of cold chemistry, which 
should ultimately lead to understanding how to manipulate reaction dynamics in realistic ultracold gases.

The paper is organized as follows. In section II we present in detail the close-coupling (CC) and MQDT formalisms for non-reactive scattering in 
collisions between two $^1\Sigma$ molecules. In section III we provide numerical illustration of the method for quasi-resonant rotational and
vibrational transitions in H$_2$-H$_2$ collisions, including both ortho and para symmetries.  Conclusions and
future directions are presented in section IV.

\section{Theory}

\subsection{Quantum close-coupling approach for molecule-molecule collisions}

The molecule-molecule scattering theory for collisions of two $^1\Sigma$ diatomic molecules has been well established and described in detail in
many prior works \cite{Taka,Green,Alex,Clary,Bala1}. Only a brief description to introduce the key 
terminologies and set the stage  
for the MQDT formalism is given here. The full close coupling (CC) \cite{Dal} methodology based on the solution of the time-independent
Schr\"odinger equation is used to solve 
the four-body scattering problem in Jacobi coordinates. 
After elimination of center-of-mass motion, the Hamiltonian for the relative motion of two H$_2$ molecules in 
space-fixed coordinates may be 
written as 
\begin{equation}
\hat{H}=-\frac{\hbar^2}{2\mu}\frac{\partial^2}{\partial R^2}+\frac{\hat{\ell}^2}{2\mu R^2}+\sum_{i=1}^2\hat{h}_{i}(r_i)+
U({\bf r_1}, {\bf r_2}, {\bf R})
\label{H4-hamiltonian}
\end{equation}
where ${\bf R}$ is the vector joining the center of mass of the two H$_2$ molecules, $\mu$ 
and $\hat{\ell}$ are the 
reduced mass and  orbital angular momentum of the two colliding H$_2$ molecules  and 
$U({\bf r_1}, {\bf r_2}, {\bf R})$ is the interaction potential. The
terms,  $\hat{h}_{i}(r_i),~i=1,2$ are the
Hamiltonians of the two isolated H$_2$ molecules:
\begin{equation}
\hat{h}_i(r_i)=-\frac{\hbar^2}{2\mu_i}\frac{\partial^2}{\partial r_i^2}+\frac{\hat{j}_i^2}{2\mu_ir_i^2}+V(r_i)
\end{equation}
where $r_i$, $\mu_i$, and $\hat{j}_i$ are the internuclear separation, reduced mass, and the rotational angular momenta of the two separated 
H$_2$ molecules.
The H$_2$-H$_2$ interaction potential  is 
expanded in terms of coupled spherical harmonics \cite{Green} 
\begin{equation}
 U({\bf r_1}, {\bf r_2}, {\bf R}) = \sum_{\lambda}A_{\lambda}(r_1,r_2,R)Y_{\lambda}(\hat{r_1},\hat{r_2},\hat{R})
\end{equation} 
with
\begin{equation}
Y_{\lambda}(\hat{r_1},\hat{r_2},\hat{R}) = \sum_{m_{\lambda}}\langle \lambda_1 m_{\lambda_1}
\lambda_2 m_{\lambda_2}|\lambda_{12} m_{\lambda_{12}}\rangle \times Y_{\lambda_1 m_{\lambda_1}}(\hat{r_1}) Y_{\lambda_2 m_{\lambda_2}}(\hat{r_2})
Y^*_{\lambda_{12} m_{\lambda_{12}}}(\hat{R})
\end{equation}
where $\lambda \equiv \{ \lambda_1,\lambda_2, \lambda_{12}\}$  and $m_{\lambda} \equiv \{  m_{\lambda_1}, m_{\lambda_2}, m_{\lambda_{12}}\}$. 
The indices
$\lambda_1$, $\lambda_2$ and $\lambda_{12}$ are non-negative integers and the sum of these three quantities must be an even integer. The
homonuclear symmetry of  H$_2$ requires that $\lambda_1$ and $\lambda_2$ must be even. The quantity in 
angular brackets of the above equation is 
a Clebsch-Gordan coefficient, and $Y_{\lambda m_{\lambda}}$ are spherical harmonics. 
The Sch\"rodinger equation is conveniently formulated by introducing the total angular momentum representation \cite{Dal}. 
The total angular momentum 
$\overrightarrow{J}=\overrightarrow{j_{12}}+\overrightarrow{\ell}$ is the vector sum of total rotational angular momentum $\overrightarrow{j_{12}}=\overrightarrow{j_1}+\overrightarrow{j_2}$ of the two molecules and orbital angular
momentum $\overrightarrow{\ell}$. 
Note that all molecules remain in singlet electronic spin states, so we suppress this notation in the following.
For collisions between two indistinguishable molecules, the total wave function $\Psi$ may be expanded in terms 
of rotational and vibrational wave functions of the two H$_2$ molecules, $\Phi_{vj\ell}^{JM\epsilon_I \epsilon_P}$, in the 
total angular momentum representation \cite{Dal}:
\begin{equation}
\Psi({\bf r_1}, {\bf r_2}, {\bf R}) = \frac{1}{R} \sum_{v,j,\ell,J,M}
F^{JM\epsilon_I \epsilon_P}_{vj\ell}(R) \Phi_{vj\ell}^{JM\epsilon_I \epsilon_P}({\bf r_1}, {\bf r_2}, {\bf R})\label{psi}
\end{equation}
where $F^{JM\epsilon_I \epsilon_P}_{vj\ell}(R)$ are the radial expansion coefficients, $v \equiv v_1,v_2$ represents the vibrational
quantum numbers and $j \equiv j_1,j_2$ specifies the rotational quantum numbers of the two 
diatomic fragments. The quantity $\epsilon_I = (-1)^{j_1+j_2+\ell}$ is the eigenvalue of the spatial inversion operator, and $\epsilon_P$
is the eigenvalue of the exchange permutation symmetry operator for two H$_2$ molecules 
(for the indistinguishable case, e.g., para-para or ortho-ortho). The explicit expression for $\Phi_{vj\ell }^{JM\epsilon_I \epsilon_P}$ 
is given in Eqs. (6), (8) and (15) of Ref.~\cite{Bala1}.  
The radial expansion coefficients  $F^{JM\epsilon_I \epsilon_P}_{vj\ell}$ are evaluated by solving the close-coupled radial equations in $R$,
\begin{eqnarray}
\left(-\frac{\hbar^2}{2\mu}\frac{d^2}{dR^2}+\frac{\hbar^2\ell(\ell+1)}{2\mu R^2}+\epsilon_{vj}-E\right) F^{JM\epsilon_I \epsilon_P}_{vj\ell} (R)
\nonumber \\ + \sum_{v',j',\ell'} {\cal U}^{JM\epsilon_I \epsilon_P}_{v j \ell,v' j' \ell'}(R)F^{JM\epsilon_I \epsilon_P}_{v'j'\ell'} (R)= 0
\label{CC_radial_eq}
\end{eqnarray}
 resulting from substitution of Eqs. (\ref{H4-hamiltonian}) and (\ref{psi}) in the time-independent Schr\"odinger equation
 $H\Psi = E\Psi$. 
Here $E$ is the total energy of the system, and we define the collision energy to be
 $E_c=E-\epsilon_{v_1j_1}+\epsilon_{v_2j_2}=E-\epsilon_{vj}$. The symbol $\epsilon_{v_i j_i} (i=1,2)$ denotes the 
 asymptotic ro-vibrational energies of the two H$_2$ molecules. Under molecule permutation the interaction potential, 
 ${\cal U}^{JM\epsilon_I \epsilon_P}_{v j \ell,v' j' \ell'}(R)$, is given by
 \begin{equation}
  {\cal U}^{JM\epsilon_I \epsilon_P}_{v j \ell,v' j' \ell'}(R) = \Delta_{vj_1j_2}\Delta_{v'j'_1j'_2}
  [{\cal U}^{JM\epsilon_I }_{v j \ell,v' j' \ell'}(R)+\epsilon_P(-1)^{j'_1+j'_2+j'_{12}+\ell'}
  {\cal U}^{JM\epsilon_I }_{v j \ell,{\bar v}' {\bar j}' \ell'}(R)],
 \end{equation} where $\bar v = v_2 v_1$ and $\bar j = j_2 j_1 j_{12}$,  $\epsilon_P=\pm 1$, and $\Delta_{vj_1j_2}=
 [2(1+\delta_{v_1v_2}\delta_{j_1j_2})]^{-1/2}$.
 The  matrix elements of the interaction potential, ${\cal U}^{JM\epsilon_I }_{v j \ell,v' j' \ell'}(R)$, are defined as
\begin{eqnarray}
{\cal U}^{JM\epsilon_I}_{v j \ell,v' j' \ell'}(R) = 
\sum_{\lambda} B^{\lambda \epsilon_I}_{vj_1j_2,v'j'_1j'_2}(R) f^{J;\lambda}_{jl,j'l'} 
\end{eqnarray}
where the radial elements $B^{\lambda \epsilon_I}_{vj_1j_2,v'j'_1j'_2}(R)$ 
are given by 
\begin{equation}
B^{\lambda \epsilon_I }_{vj_1j_2,v'j'_1j'_2}(R) =
\int_0^{\infty} \int_0^{\infty} \chi_{vj_1j_2}^{\epsilon_I}(r_1,r_2) 
A_{\lambda}^{\epsilon_I}(r_1,r_2,R)  \chi_{v'j'_1j'_2}^{\epsilon_I}(r_1,r_2) dr_1 dr_2  ,
\end{equation}
and the  function $f^{J;\lambda}_{jl,j'l'}$
is given in terms of $3-j$, $6-j$, and $9-j$ symbols:
\begin{multline}
f^{J;\lambda}_{jl,j’l’} = 
(4\pi)^{-3/2} (-1)^{j_1+j_2+j'_{12}+J}[\lambda,j,l,j',l',\lambda_{12}]^{1/2} \\ 
\left( \begin{array}{ccc} j_1 & j'_1 & \lambda_1 \\ 0 & 0 & 0 \end{array} \right)
\left( \begin{array}{ccc} j_2 & j'_2 & \lambda_2 \\ 0 & 0 & 0 \end{array} \right)
\left( \begin{array}{ccc} l & l' & \lambda_{12} \\ 0 & 0 & 0 \end{array} \right) \\ 
\left\{ \begin{array}{ccc} l & l' & \lambda_{12} \\ j'_{12} & j_{12} & J \end{array} \right\}
\left\{ \begin{array}{ccc} j'_{12} & j'_2 & j'_1 
\\ j_{12} & j_2 & j_1 \\ \lambda_{12} & \lambda_2 & \lambda_1  \end{array} \right\}, 
\end{multline}

with the notation
\begin{eqnarray}
[x_1, x_2, ... , x_n] = (2x_1+1) (2x_2+1) ... (2x_n+1).
\end{eqnarray}

 In the coupled-channel formalism, either the wave function ${\bf F}(R)$ and its derivative
${\bf F}'(R)$ or the log-derivative matrix ${\bf Y=F'F^{-1}}$ is propagated from a point in the classically forbidden 
region near the origin, $R\sim0$, to where the interaction potential becomes negligible, $R_{\infty}$.
In the present case, the CC equation for each value of $R$ is solved by propagating the log-derivative matrix ${\bf Y}$ by 
following the methods of 
Johnson and Manolopoulos \cite{John,Mana}.  The scattering matrix ${\bf S}$ for 
specific values of $J$, $\epsilon_I$ and $\epsilon_P$ is evaluated by matching the ${\bf Y}$ matrix to known 
asymptotic solutions of the CC equations at  $R_{\infty}$. The boundary condition is
\begin{equation}
 ({\bf Y}{\bf J}-{\bf J}')=({\bf Y}{\bf N}-{\bf N}'){\bf K}.
\end{equation}
The matrices ${\bf J}$ (not to be confused with the total angular momentum) and ${\bf N}$ are  diagonal matrices of asymptotic functions.
For convenience, the total number of coupled-channels $N$ is partitioned into 
$N_o$ open channels (with $E>0$) and $N_c$ closed channels (with $E \le 0$) such that $N=N_o+N_c$. 
For the open channels $N_o$ these functions are known as Riccati-Bessel functions, and for
the closed channels $N_c$ they are modified spherical Bessel functions of the first and third kinds
\cite{ABRAMOWITZ}. ${\bf J}'$ and ${\bf N}'$ are the
derivative matrices of ${\bf J}$ and ${\bf N}$, respectively. For an $N$ channel problem the scattering 
${\bf S}$ matrix is easily calculated by considering 
only the open-open sub-block of ${\bf K}$ matrix by the following expression
\begin{equation}
{\bf S} = (1+i{\bf K}_{oo})^{-1}(1-i{\bf K}_{oo}). 
\end{equation}
Finally, the state-to-state cross section is obtained from the ${\bf S}$ matrix.
For indistinguishable molecule collisions  one must symmetrize the cross-section with the statistically weighted sum of the exchange-permutation 
symmetry components. Explicit expressions for  state-to-state cross section with and without exchange symmetry have been given in prior 
publications \cite{Bala1,Bala2}. For completeness, and for the ease of comparisons with MQDT results, the expressions for
the symmetrized cross sections are reproduced below: 
\begin{equation}
 \sigma_{v_1j_1v_2j_2\rightarrow v'_1j'_1v'_2j'_2}(E_c) = W^+\sigma^{\epsilon_P=+1} +  W^-\sigma^{\epsilon_P=-1}
\end{equation}
with
\begin{eqnarray}
 \sigma^{\epsilon_P} &=&  \frac{\pi(1+\delta_{v_1v_2}\delta_{j_1j_2})(1+\delta_{v'_1v'_2}\delta_{j'_1j'_2})}{(2j_1+1)(2j_2+1)k^2}
 \\
 &\times& \sum_{j_{12}j'_{12}\ell\ell' J\epsilon_I} (2J+1)|\delta_{vj\ell,v'j'\ell'}-S^{J\epsilon_I \epsilon_P}_{vj\ell,v'j'\ell'}(E_C)|^2
\label{Cross_ident}
\end{eqnarray}
where $k^2 = 2 \mu E_c /\hbar^2$. In the case of collisions of two ortho-H$_2$ molecules having nuclear spin $I=1$
and weight factors $W^+=2/3$ and $W^-=1/3$,
one must consider both exchange permutation symmetries $\epsilon_P=\pm 1$ for the calculation of state-to-state cross sections.
For collisions
between two para-H$_2$ molecules with nuclear spin $I=0$ and weight factors $W^+=1$ and $W^-=0$, only one exchange-permutation symmetry 
$\epsilon_P=+1$ is required for evaluating the cross-section. To describe the state-to-state cross-section between two H$_2$
molecules, we use the 
term 
``combined molecular state", CMS, which denotes the combined ro-vibrational quantum numbers of the two molecules. In this notation, the collision 
of the first
H$_2$  molecule having the ro-vibrational state $(v_1,j_1)$ with the second H$_2$ molecule in state ($v_2,j_2$)  is denoted by a unique term 
($v_1,j_1,v_2,j_2$). This CMS is the quantum state which characterizes 
the molecule-molecule system before or after the collision.

It should be emphasized that the CC method described here is a numerically exact calculation that incorporates the complete four-body 
physics of the collision complex, provided that sufficiently many channels are included in the calculation (which is certainly possible for 
light molecules such as H$_2$). This method does, however, require the complete calculation to be performed separately for each collision energy 
of interest.  The number of such calculations may be large, say in the case where cross sections vary with energy due to resonances or
(at ultracold temperature) due to the Wigner threshold laws.  Restricting this requirement of calculations at many energies is a 
main accomplishment of the MQDT method, to which we now turn.

\subsection{The MQDT Formalism}
The MQDT formalism modifies the scattering calculations in several ways.  First, it acknowledges that, beyond a
certain interparticle spacing $R_m$, the scattering channels become independent from one another, and their wave functions can be 
constructed in each channel individually. This leads to a reduction in computational time since the number of arithmetic operations is proportional
to $N^3$ for the CC calculation.
Whereas, this number is only proportional to $N$ for the MQDT calculation. Second, it notes that this distance $R_m$ can often be chosen small 
enough that all channels 
are ``locally open,'' meaning that the kinetic energy at $R_m$ is positive in each channel.  In this circumstance, boundary conditions in closed 
channels need not yet be applied, and the wave function at $R_m$ will not have the sensitive energy dependence required near resonances.  

Third, the asymptotic wave functions to which one matches the short-range wave function are themselves chosen to exhibit weak energy dependence, 
so that the resulting short-range $K$-matrix, $\bf K^{\rm sr}$, is only weakly dependent on energy and magnetic field.  This allows for efficient 
calculations over a wide range of energy and field.
Features such as resonances and Wigner threshold laws are then recovered at a later stage via relatively simple algebraic procedures.
This method has proven useful and economical in molecular scattering \cite{Mies1} and in ultracold collisions \cite{Burke}.  Here we describe 
its application to the H$_2$-H$_2$ cold collision problem.

$\bf{K}^{\rm sr}$ is defined by writing the matrix wavefunction, $\bf{M}$, in terms of MQDT reference functions, $\hat{f}$ and $\hat{g}$,
\begin{equation}
 M_{ij} = \hat{f}_i\delta_{ij} - \hat{g}_i K^{sr}_{ij} \quad \text{for } R\ge R_m
\end{equation}
where $\bf{M}$ is a $N\times N$ matrix that contains $N$ wavefunctions with physical boundary conditions at the origin.
${\bf K}^{\rm sr}$ is obtained by matching the log-derivative of $\bf{M}$ to the log-derivative matrix
${\bf Y}$ at $R_m$
\begin{eqnarray}
  \bf{M}^{'}\bf{M}^{-1}=\bf{Y} \\
  {\bf K}^{\rm sr} = ({\bf Y}\hat{g}-\hat{g'})^{-1}({\bf Y}\hat{f}-\hat{f'}).
\end{eqnarray}

To achieve a weakly energy and field dependent $\bf K^{\rm sr}$, we let $\hat{f}$ and $\hat{g}$ have WKB-like boundary 
conditions well within the classically allowed region at $R=R_x \le R_m$ \cite{Mies1,Rau},
\begin{eqnarray}
\hat{f_i} (R) = \frac{1}{\sqrt{ k_i(R)}} {\rm sin} \left( \int_{R_x}^Rk_i(R') dR'+\phi_i\right) \hspace{0.2cm} {\rm at}  \hspace{0.2cm} R=R_x 
\label{f_hat}\\
\hat{g_i} (R) = -\frac{1}{\sqrt {k_i(R)}} {\rm cos} \left( \int_{R_x}^Rk_i(R') dR'+\phi_i\right) \hspace{0.2cm} {\rm at}  \hspace{0.2cm} R=R_x 
\label{g_hat}
\end{eqnarray}
where $\phi_i$ denotes an energy independent phase described by Ruzic et al. \cite{Ruzic13_PRA}. Here, $k_i(R)=\sqrt{\frac{2 \mu}{\hbar^ 2} 
\left(E-\epsilon_{v_i j_i}-V_{ii}(R)\right)}$, and the derivatives of
$\hat{f}$ and $\hat{g}$ at $R_x$ are defined by the full, radial derivatives of equations (\ref{f_hat}) and (\ref{g_hat}).

One obtains $\hat{f}$ and $\hat{g}$ at all $R$ by solving a 1-D Schr\"odinger equation
\begin{eqnarray}
 \left(-\frac{\hbar^2}{2\mu}\frac{d^2}{dR^2}+\frac{\hbar^2\ell_i(\ell_i+1)}{2\mu R^2}+V^{lr}+\epsilon_{v_i j_i}-E\right) {\hat{f_i} 
 \choose \hat{g_i}}= 0 
\label{MQDT_reffun}
 \end{eqnarray}
subject to the boundary conditions (\ref{f_hat}) and (\ref{g_hat}). For H$_2$-H$_2$ scattering, beyond the strong interaction region, one only needs to deal with the weak, attractive van der Waals forces. Hence, the MQDT reference functions can be obtained by choosing a long-range expansion for the reference potential,
$V^{\rm lr}=-\frac{C_6}{R^6}-\frac{C_8}{R^8}-\frac{C_{10}}{R^{10}}$.

The matrix ${\bf K}^{\rm sr}$ and the linearly independent solutions, $\hat{f}$ and $\hat{g}$, carry all the information required to obtain the scattering 
observables. To obtain the physical scattering matrix,  ${\bf S}^{\rm phys}$, four MQDT parameters ${\cal A}$, ${\cal G}$, $\eta$
and $\gamma$ are required in each channel \cite{Ruzic13_PRA}. These four quantities correctly describe the asymptotic behavior of the reference wave 
functions $\hat{f}$ and $\hat{g}$. Explicit expressions for these parameters are given in Eqs (12a) to (12d) in \cite{Ruzic13_PRA}. 

By partitioning ${\bf K}^{\rm sr}$ into energetically open (o) and closed (c) channels, we eliminate the unphysical growth inherent in $\bf M$ by the following transformation 
\begin{equation}
\label{reduction}
 \tilde {\bf K} = {\bf K}^{\rm sr}_{oo}-{\bf K}^{\rm sr}_{oc}\left({\rm cot}\gamma+{\bf K}^{\rm sr}_{cc}\right)^{-1}{\bf K}^{\rm sr}_{co}
\end{equation}
where ${\rm cot}\gamma$ is a diagonal matrix of dimension $N_c\times N_c$. 
Hence, $\tilde{\bf K}$, represents the $N_o$ wavefunctions with physical boundary conditions both at the origin and asymptotically. 
Roots of det$(\bf{K}_{\rm cc}+\cot\gamma)$ approximate the locations of resonances in the cross section.

In order to relate $\tilde{\bf K}$ to ${\bf S}^{\rm phys}$, another set of energy-normalized, 
linearly independent solutions is required. For each energetically open channel, the reference functions $f$ and $g$ are defined as
\begin{eqnarray}
 f_i(R) \xrightarrow{R\rightarrow\infty} k_i^{-1/2}{\rm sin}(k_i R-\ell_i\pi/2+\eta_i) \\
 g_i(R) \xrightarrow{R\rightarrow\infty} -k_i^{-1/2}{\rm cos}(k_i R-\ell_i\pi/2+\eta_i)
 \end{eqnarray}
These functions are related to $\hat{f}$ and $\hat{g}$ through the following expressions,
\begin{eqnarray}
 f_i(R) = {{\cal A}_i}^{1/2}\hat{f_i}(R)\label{rela1}\\
 g_i(R) = {{\cal A}_i}^{-1/2}{\cal G}_i\hat{f_i}(R)+{{\cal A}_i}^{-1/2}\hat{g_i}(R)\label{rela2}.
\end{eqnarray}
Hence, $\bf S^{\rm phys}$ is obtained by the following series of simple transformations
\begin{eqnarray}
 {\bf K} = {\cal A}^{1/2}{\tilde {\bf K}}\left(I+{\cal G}{\tilde {\bf K}}\right)^{-1}{\cal A}^{1/2} \\
{\bf S}^{\rm phys} = e^{i\eta}\left(I+i{\bf K}\right)\left(I-i{\bf K}\right)^{-1}e^{i\eta}
\end{eqnarray}
where $\cal A$ and $\cal G$ and $\eta$ are diagonal matrices of order $N_o\times N_o$ and $I$ is the identity matrix.


\section{Results}

Our main goal in this article is to demonstrate the power of the MQDT method for ultracold, non-resonant and quasi-resonant molecular scattering. 
To this end we 
wish to establish two criteria for H$_2$.  First, that the separation between long and short-range physics is reasonable; and second, 
that the energy-dependent scattering may be easily described via an essentially energy-independent short-range wave function. We will also
examine the sensitivity of results to the choice of the reference potential. To address these 
criteria, we will focus on H$_2$ collisions in which energy is nontrivially transferred among vibrational and rotational degrees of freedom between
the two molecules.


\subsection{Quasi-resonant scattering: convergence with respect to matching distance $R_m$}

To establish the use of MQDT as a reasonable separation between short- and long-range behavior, we choose a problem where nontrivial energy 
exchange occurs in the short-range physics.  The specific example is one of ``quasi-resonant'' energy transfer in para-para H$_2$ scattering,
whereby two units of rotational angular momentum are transferred from a vibrationless molecule to a vibrating molecule \cite{Bala3,Bala2}.  
\begin{figure}[tbp]
\centering
\begin{tabular}{cc}
\includegraphics[width=0.45\columnwidth]{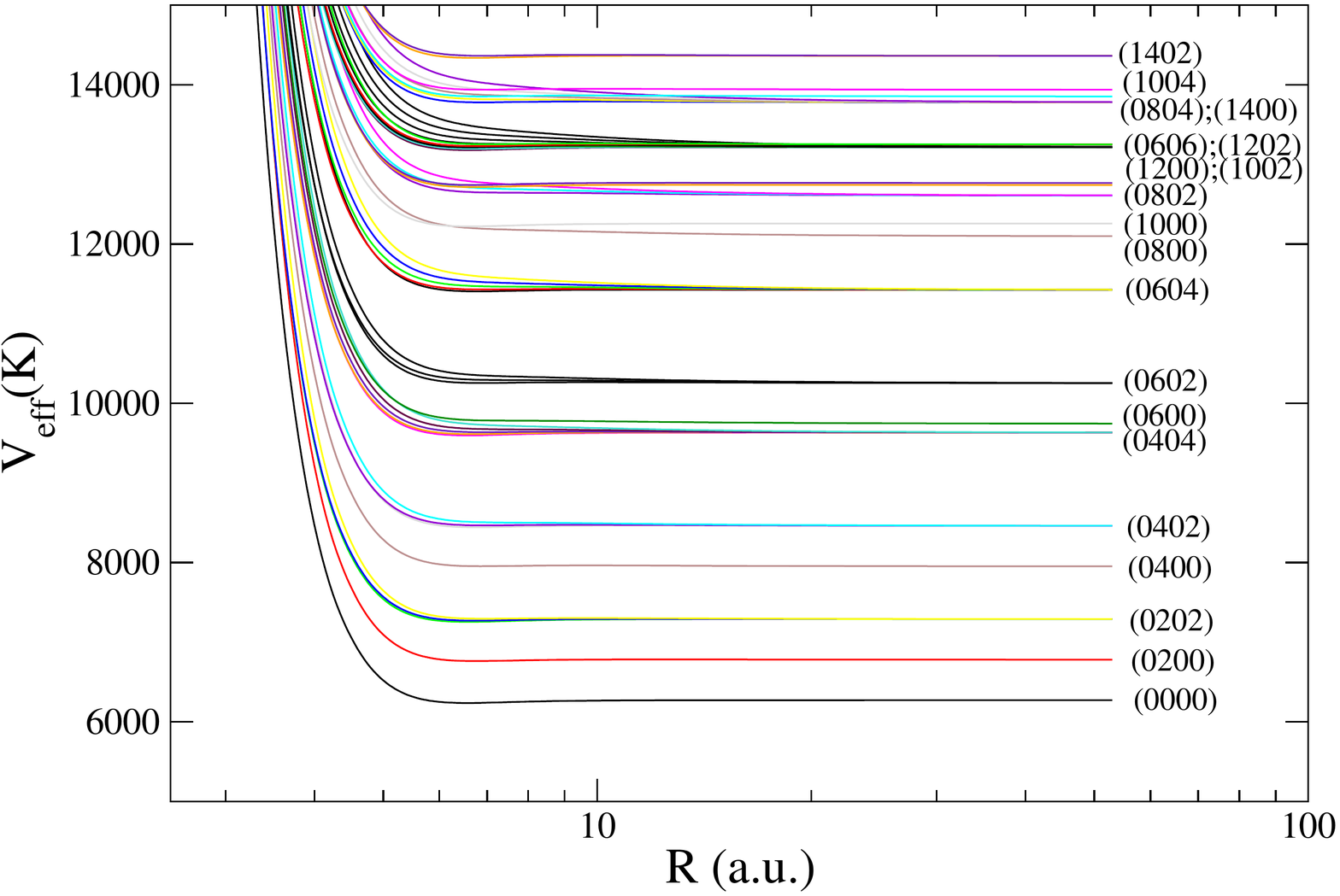}
\includegraphics[width=0.45\columnwidth]{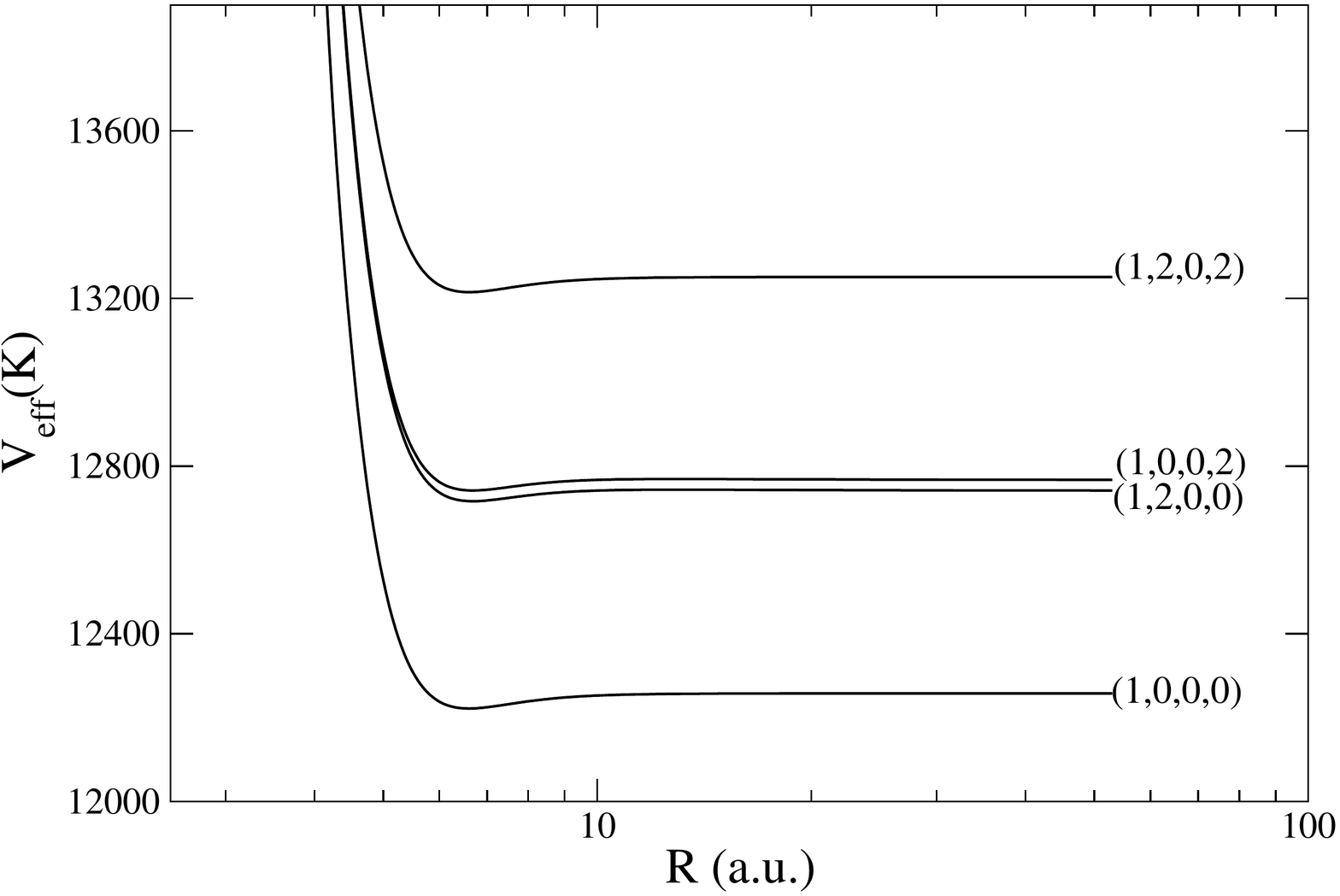}
\end{tabular}
\caption{Effective potential energy curves for para-H$_2$ scattering as defined in eq.(\ref{effective-V}). 
The labels for nearly degenerate curves are separated by a colon,
where the highest lying threshold is on the right. The panel on the right shows the CMSs included in the basis set.}
\label{effective_PP}
\end{figure}
To illustrate this process, diabatic potential energy curves for two interacting para-H$_2$ molecules versus the intermolecular separation
$R$ are
shown in Fig.~\ref{effective_PP}. These are effective potentials, defined as 
\begin{equation}
V_{eff} (R) = \epsilon_{vj} + {\cal U}^{JM\epsilon_I \epsilon_P}_{v j \ell,v j \ell}(R) + \frac{\hbar^2\ell(\ell+1)}{2\mu R^2}.
\label{effective-V}
\end{equation}

The particular 
quasi-resonant process of interest takes the initial channel $(v_1,j_1,v_2,j_2)$ $=(1,0,0,2)$ to $(1,2,0,0)$.  The name 
quasi-resonant pertains to the fact that the thresholds of these channels are nearly degenerate, as seen in Fig.~\ref{effective_PP}.  Specifically, 
their energy separation, 25.45 K, is comparable to the well depth of the isotropic part of the H$_2$-H$_2$ interaction $\sim$ 31.7 K. 
The slightly different centrifugal
distortion of the vibrational levels $v=0$ and $v=1$ is responsible for this small energy gap between the two CMSs. In this process the 
total rotational angular momentum is conserved by the collision. These kinds of  transitions which have a small internal energy
and  internal rotational angular momentum gaps are found to be very efficient and highly state-selective and have 
been referred to as quasi-resonant rotation-rotation (QRRR) transfer \cite{Bala3}.
Note that alternative final states are also possible, but the one we have
selected is known to be the dominant one.  Indeed, as illustrated in \cite{Samatha,Samantha1}, 
to accurately describe such quasi-resonant energy transfer, one does not need
to couple any other $v,~j$ levels in the basis set. One can restrict the basis set in the CC calculations to just those involved 
in the quasi-resonant transition yet still get results comparable to those from  a larger basis set that includes many other CMSs. Thus,
for the purpose of  simplicity, we have resorted to a small basis (1,0,0,0),(1,0,0,2),(1,2,0,0) and (1,2,0,2) (as shown in the right panel of
Fig.~\ref{effective_PP}) that primarily includes the quasi-resonant
channels in the CC calculations.

Using the restricted basis set described above, 
we have computed scattering cross sections for this process using the full CC calculation and the MQDT 
formalism. In the full CC calculation asymptotic matching to free-particle wave functions is carried out at $R_{\infty}$= 100 $a_0$. 
In the MQDT approach, reference functions are determined
from both the isotropic parts of the diagonal elements of the long-range diabatic potential coupling matrix, 
${\cal U}^{JM\epsilon_I \epsilon_P}_{v j \ell,v j \ell}(R)$, and also the long-range approximation 
for the reference potential, $V^{\rm lr}=-\frac{C_6}{R^6}-\frac{C_8}{R^8}-\frac{C_{10}}{R^{10}}$. As 
discussed in the next subsection, we find that more accurate results are obtained
 when the long-range part of the 
diabatic potential curves are employed.

\begin{figure}[tbp]
\centering
\begin{tabular}{cc}
\includegraphics[width=0.45\columnwidth]{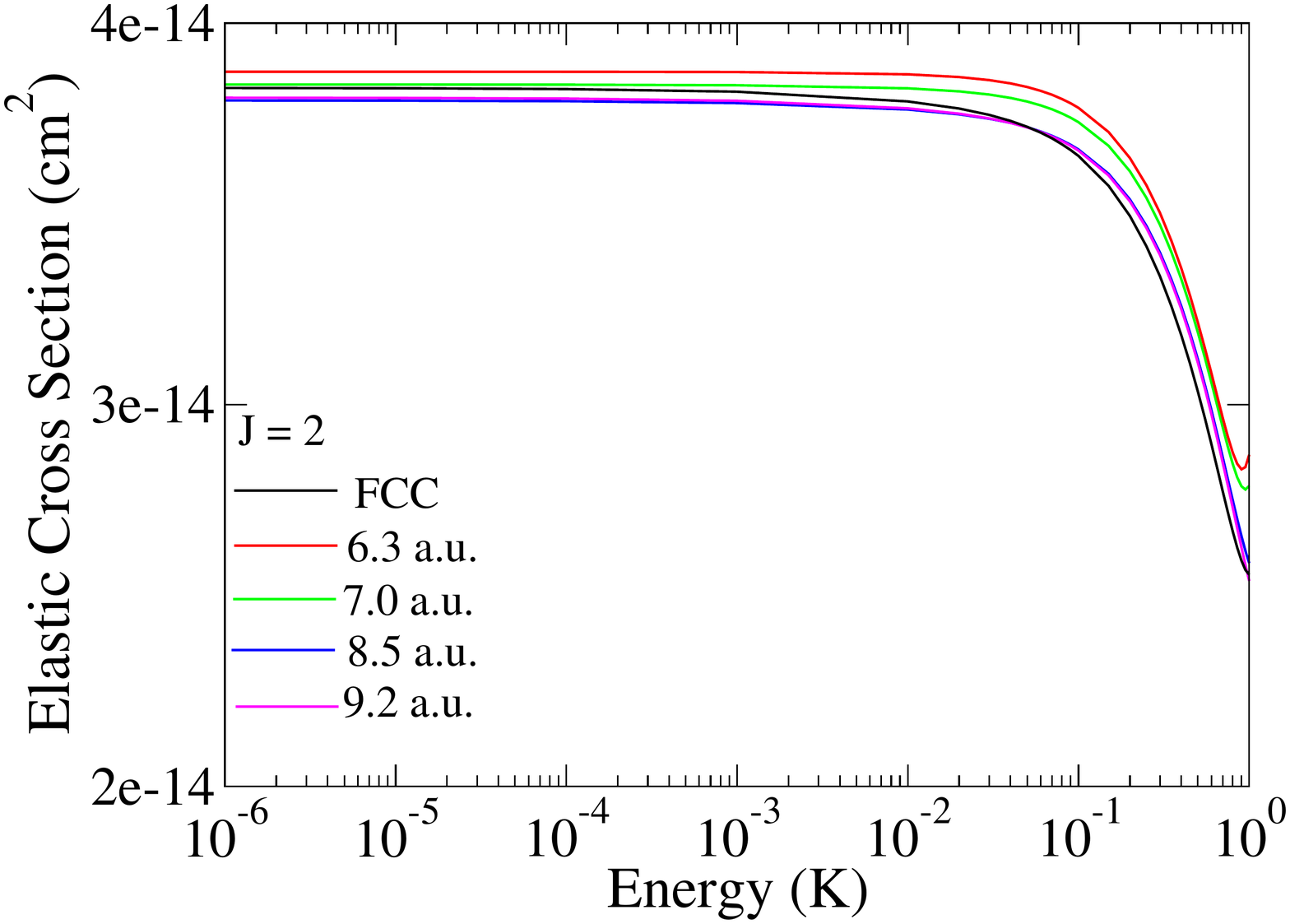}
\includegraphics[width=0.45\columnwidth]{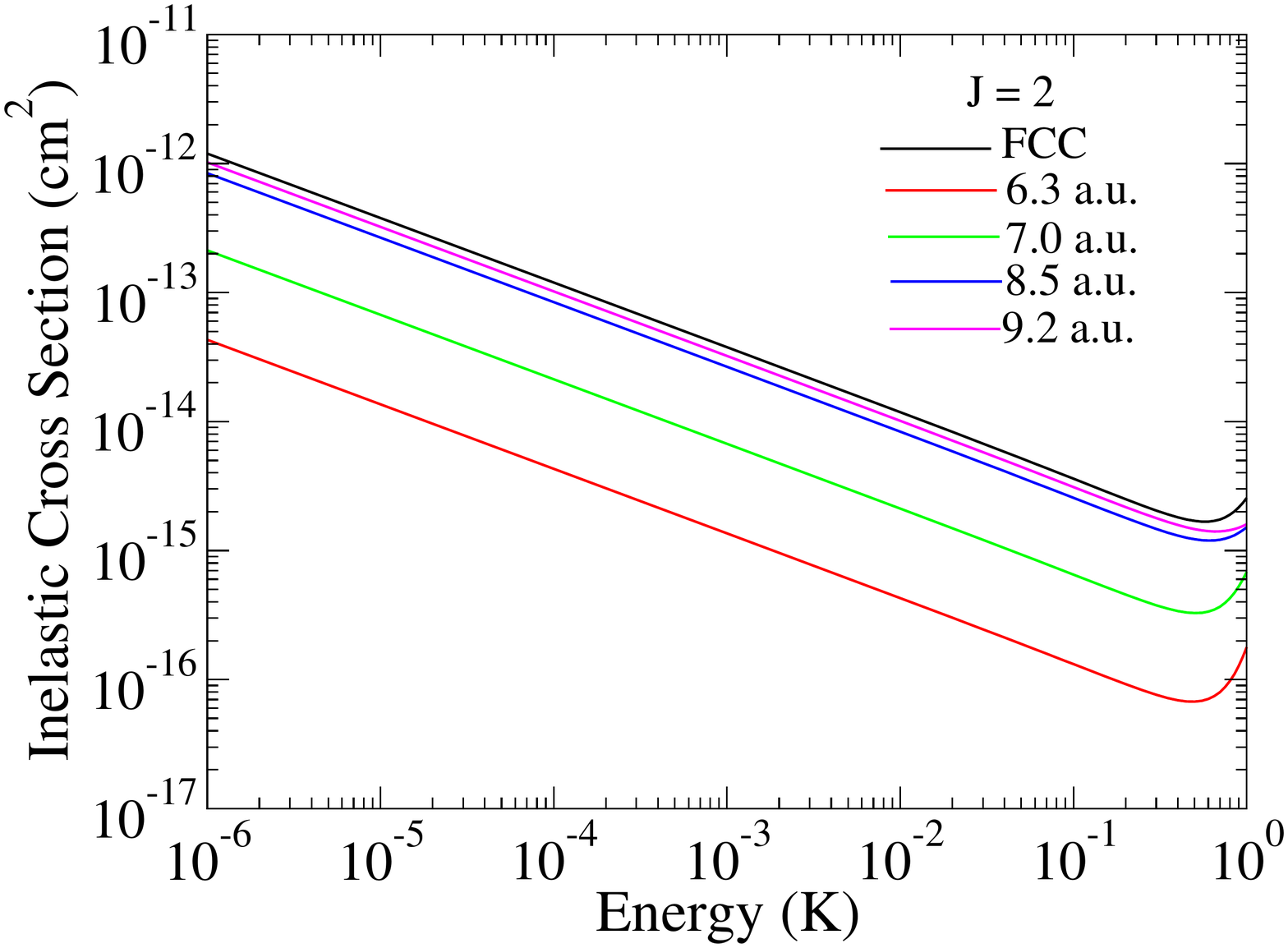}
\end{tabular}
\caption{ Elastic (left) and inelastic (right) cross sections for (1,0,0,2) to (1,2,0,0) quasi-resonant 
scattering in para-H$_2$.  The black curves 
correspond to the full CC calculation, while the other curves correspond to the MQDT result with different matching 
radii, $R_m$.} 
\label{QRconverge}
\end{figure}

First, we establish convergence of elastic and inelastic cross sections as a function of 
the short-range matching distance $R_m$. Results of these studies are shown in Figure \ref{QRconverge}, for elastic (left panel) and inelastic QRRR (right panel) scattering.  The QRRR transition is the dominant inelastic channel
that corresponds to total angular momentum $J=2$ and $s$-wave scattering in the incident channel.  The 
solid black curve refers to results from the full close-coupling calculation, while the other curves correspond 
to the MQDT results for various values of the 
matching radius $R_m$. The convergence with respect to $R_m$ is excellent and occurs 
as soon as $R_m$ exceeds the region of the van der Waals well.  The MQDT calculations are converged and nearly
quantitatively reproduce the results from the full CC calculation for a matching distance of $R_m$=9.2 $a.u$.  
Note that the van der Waals length, $r_{\rm vdw}=\left(\frac{2\mu C_6}{\hbar^2}\right)^{1/4}$ for H$_2$-H$_2$ is 14.5 $a_0$.
The agreement is
also nearly quantitative for elastic scattering cross sections shown in the left panel. Although, for 
elastic scattering, the results are less sensitive to the matching distance.
It should be emphasized that these convergence tests involve a single 
short-range ${\bf K}^{\rm sr}$ matrix computed at an initial collision energy
of 1 $\mu$K, which is able to capture the dynamics at other collision energies in 1 $\mu$K-1 K range. This is an
important aspect of MQDT calculations as discussed in more detail below.

\subsection{Choice of reference potential and energy independent parameters}

A key aspect of MQDT is to simplify the calculation and description of scattering, especially at ultracold temperatures.
We have already demonstrated this for the para-para case where an energy independent short
range  ${\bf K}^{\rm sr}$ is able to reproduce cross sections over a wide range of energies when resonances are
absent. 
An equally important issue is the choice of reference potential for the evaluation of MQDT
reference functions. While for atom-atom scattering the obvious choice is the long-range expansion, for complex 
systems such as the present case, a more 
accurate choice given by the effective potential of Eq.(\ref{effective-V}) may be adopted. 
To explore these issues we  consider similar quasi-resonant energy transfer in
ortho-ortho and ortho-para collisions. Further, we extend the energy range of these 
calculations to 10 K to capture a $d$-wave shape resonance reported in an earlier work \cite{Samatha}.

\begin{figure}[tbp]
\centering
\begin{tabular}{cc}
\includegraphics[width=0.45\columnwidth]{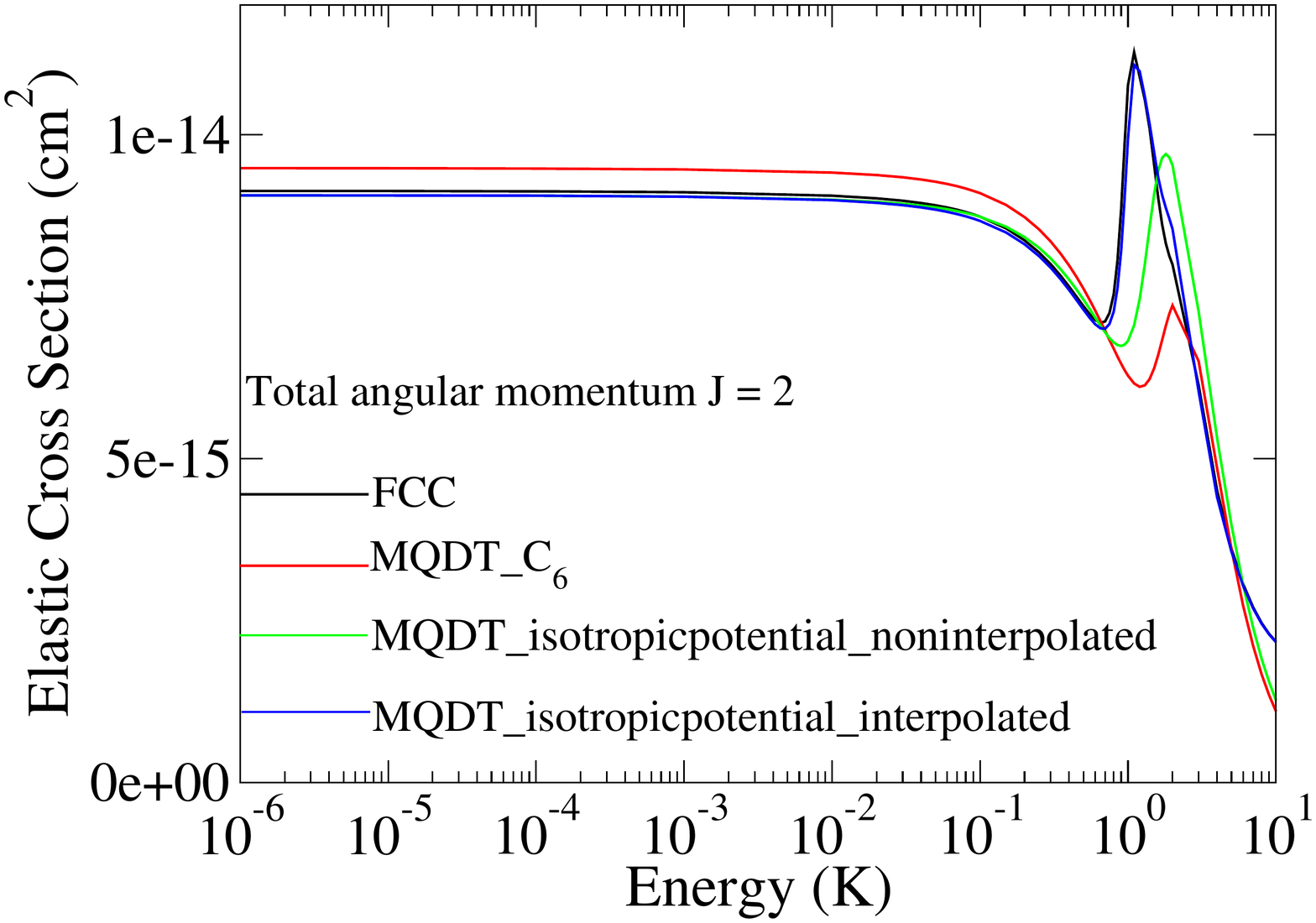}
\includegraphics[width=0.45\columnwidth]{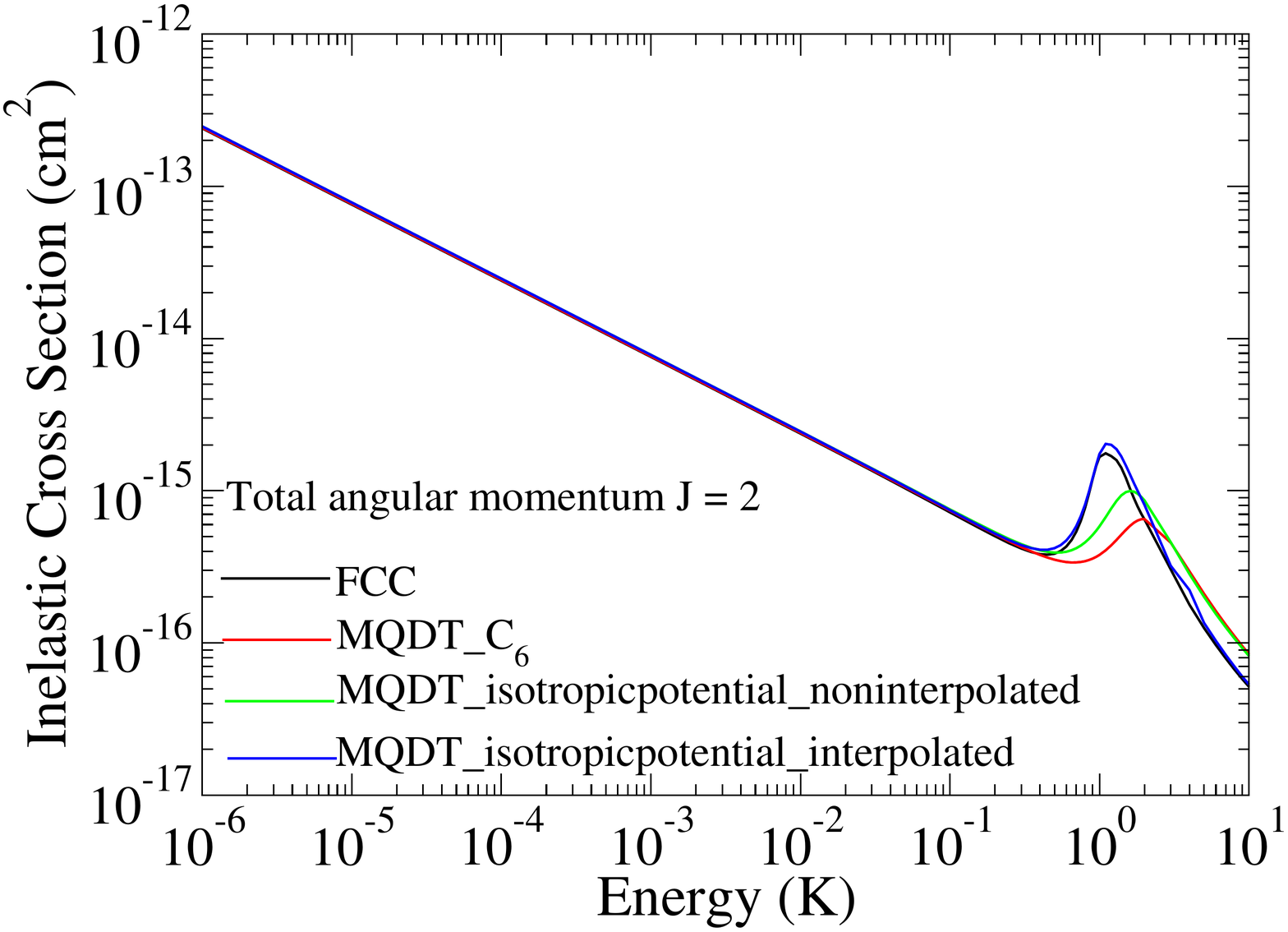}
\end{tabular}
\caption{This figure shows the elastic cross section for H$_2(v=1,j=1)$+H$_2(v=0,j=3)$ 
collisions (left panel) and the inelastic cross section for H$_2(v=1,j=1)$+H$_2(v=0,j=3)\to $ H$_2(v=1,j=3)$+H$_2(v=0,j=1)$
quasi-resonant process (right panel).}
\label{OOscatter}
\end{figure}

Similar to the case of para-H$_2$, here we consider a QRRR process in which
an ortho-H$_2$($v=1,j=1$) molecule hits  another vibrationless 
ortho-H$_2$($v=0,j=3$) 
molecule and takes away two units of rotational angular momentum. The process is described as $(v_1,j_1,v_2,j_2)$ 
$=(1,1,0,3)$ $\rightarrow$ $(1,3,0,1)$. In this case the energy difference between the two threshold channels is $45.5$ K. Full CC calculations
have been reported for this process previously \cite{Samatha,Samantha1}, but we show results for the positive exchange symmetry to 
compare with MQDT
 in Fig.\ref{OOscatter}. The elastic cross section is calculated according to equation (\ref{Cross_ident}). Only cross-sections 
 for total angular momentum $J=2$ that include s-wave scattering in the incident channel are shown. The solid black curve denotes the full CC 
 result for both elastic (left panel) and inelastic (right panel) collisions  obtained by matching to free particle wave functions at 
$R_{\infty}$= 100$ a_0$.

MQDT results are obtained using a short-range matching distance of 9.5 $a_0$. 
The different curves for MQDT refer to different choices for the reference potential. The red curve 
corresponds to using a simplified potential $V^{\rm lr} = -C_6/R^6$, while the green curve corresponds to the diagonal elements of the 
isotropic part of diabatic coupling matrix discussed above. Both results correspond to a single
${\bf K}^{\rm sr}$ matrix evaluated  at 1 $\mu$K  for the entire energy regime.  
Both reference potentials are able to identify the $d$-wave resonance near 1 K, but do not find its 
energy position particularly accurately.  In addition, the very simplest reference potential struggles to reproduce 
the non-resonant elastic cross section near threshold since this potential does not deliver an accurate phase shift in the 
region $R_m < R < R_{\infty}$.
The blue curves in Figure \ref{OOscatter}  are also obtained using the diabatic potential coupling matrix for the
reference potential, but here  ${\bf K}^{\rm sr}$ matrix has been evaluated at various 
energy values in the $1\mu$K-10 K range followed by interpolation in a fine energy grid. This grid consists of three different regions:
(i) in the ultralow energy range, $E_c=1\mu$K - 100 mK,  ${\bf K}^{\rm sr}$ matrix is evaluated at $1\mu$K, 1 mK and 
100 mK; (ii) at low energies in the  range $E_c$=200 mK-1 K,  ${\bf K}^{\rm sr}$ is evaluated at 9 points with 100 mK separation;
(iii) from 1 K - 10 K, an energy spacing of 1 K was employed.  
In this case the MQDT reproduces the full CC result quite well. MQDT identifies the resonance position and lineshape and 
matches the background cross sections at the percent level.
\begin{figure}[tbp]
\centering
\includegraphics[width=0.6\columnwidth]{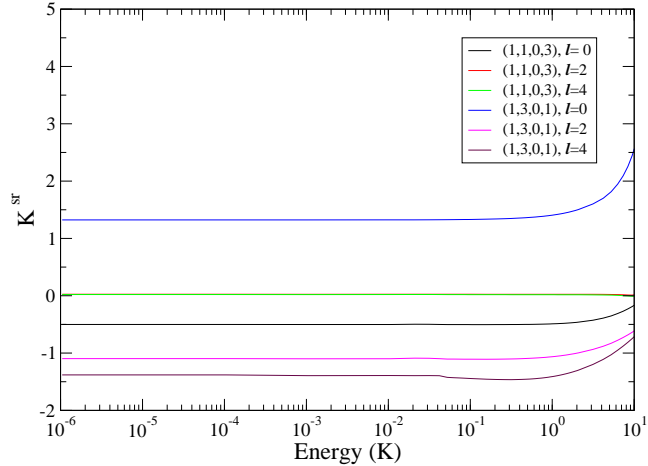}
\caption{Diagonal elements of the short-range K-matrix corresponding to the quasi-resonant transition [(1,3,0,1) $\to$ (1,1,0,3)]
as a function of the collision energy. The isotropic part of the diabatic potential matrix elements 
is used for the reference potential and the matching radius
$R_m=9.5~a_0$.}
\label{Short-range}
\end{figure}

The ability to interpolate the short-range $K$-matrix and still get quantitative results stems from the 
smoothness of this quantity in energy, as shown in Fig. \ref{Short-range}. 
Only elements corresponding to the 
quasi-resonant channels (1,1,0,3 and 1,3,0,1) are shown. Although the partial waves, $\ell =2$ and 4 are present in both the initial and final
channels, the dominant contribution comes from $\ell=0$ for $E_c=1\mu$K$-$ 1 K. It is clear from Fig.\ref{Short-range}
that up to 1 K the short-range K-matrix is independent of energy, but it becomes a smooth function of energy 
beyond 1 K. Thus, an energy independent short-range K-matrix evaluated at 1$\mu$K cannot be expected to adequately
reproduce a shape resonance near 1 K. 

Also, we note that the  well depth of the isotropic part of the
diabatic potential is only $\sim$ 31.7 K. When the collision  energy becomes a 
significant fraction of the potential well depth  the ${\bf K}^{\rm sr}$ matrix becomes 
a strong function of energy, and its energy dependence needs to be taken into account. Thus, 
for systems such as H$_2$+H$_2$ characterized by a relatively shallow van der Waals well, the energy
dependence of the short-range K-matrix becomes important in describing the 
resonances supported by the van der Waals well. 
For other systems with deeper potential wells, one may be able to use an energy 
independent ${\bf K}^{\rm sr}$ matrix over a wider range of scattering energies.
It is also possible that the MQDT treatment will need modification to handle resonances already present in ${\bf K}^{\rm sr}$ \cite{Hud2}.

Next, we demonstrate the ability of the MQDT method to describe physics driven by vibrational, rather than rotational, dynamics. Specifically, 
we consider  a second type of H$_2$ scattering, wherein a vibrationless
ortho-H$_2$ hits a para-H$_2$ molecule that carries 
one quantum of vibration, transferring this vibration to the ortho molecule \cite{Samantha1}. 
In our notation, para-H$_2$($v=1,j=0$)+ortho-H$_2$($v=0,j=1$) 
$\rightarrow$ para-H$_2$($v=0,j=0$) +ortho-H$_2$($v=1,j=1$) i.e, $(1,0,0,1)$ $\rightarrow$ $(0,0,1,1)$. Despite a small energy gap (8.5 K) 
between the initial and final states, one observes a surprisingly small inelastic cross section. This is because, due to the homonuclear symmetry 
of the H$_2$ molecules, the transfer of rotational energy between $j=0$ and $j=1$ states is symmetry forbidden. Hence the entire
process is driven by 
vibrational energy transfer, which is generally less efficient than rotational energy transfer \cite{Bala3}.

\begin{figure}[tbp]
\centering
\begin{tabular}{cc}
\includegraphics[width=0.4\columnwidth]{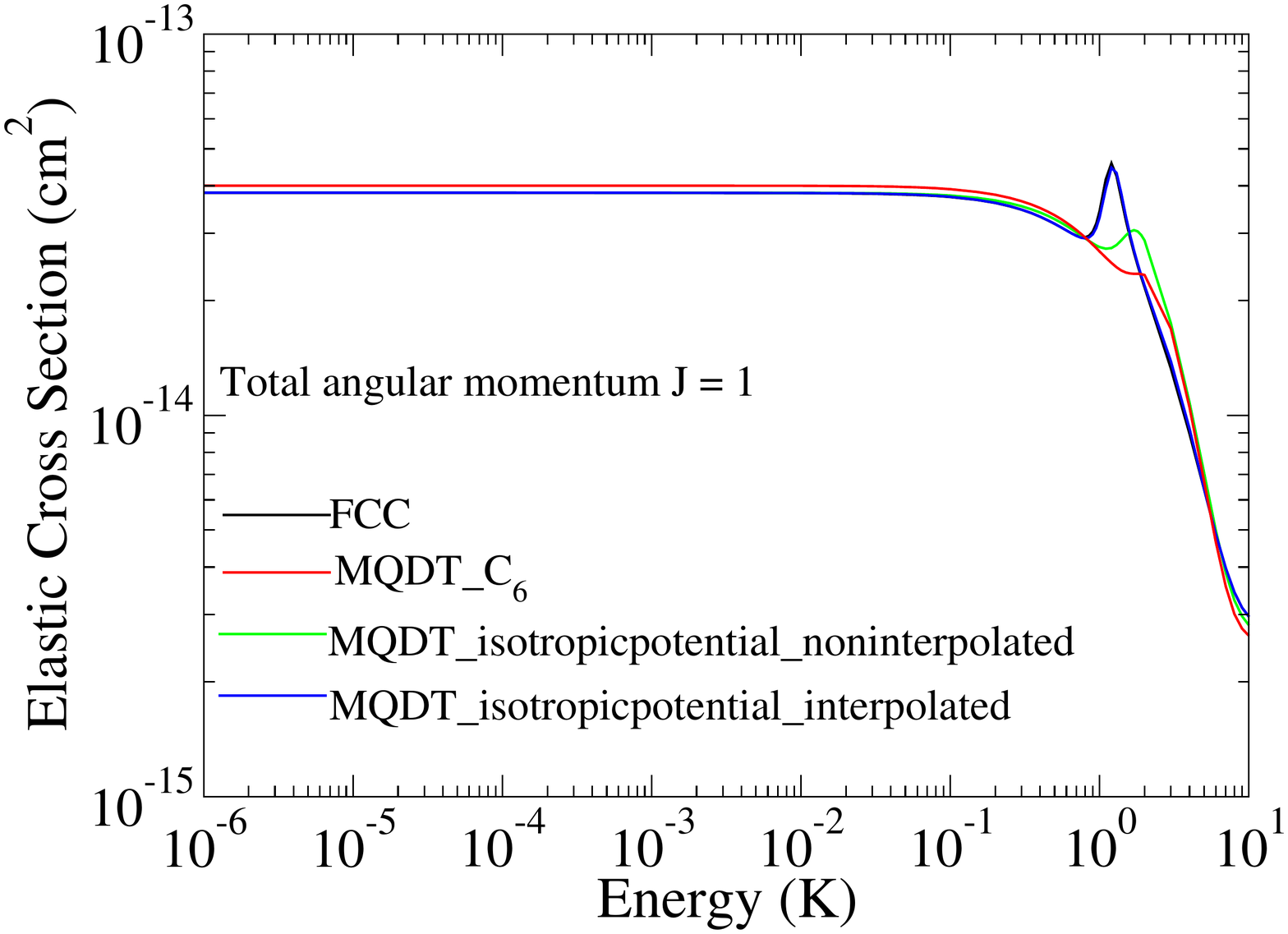}
\includegraphics[width=0.4\columnwidth]{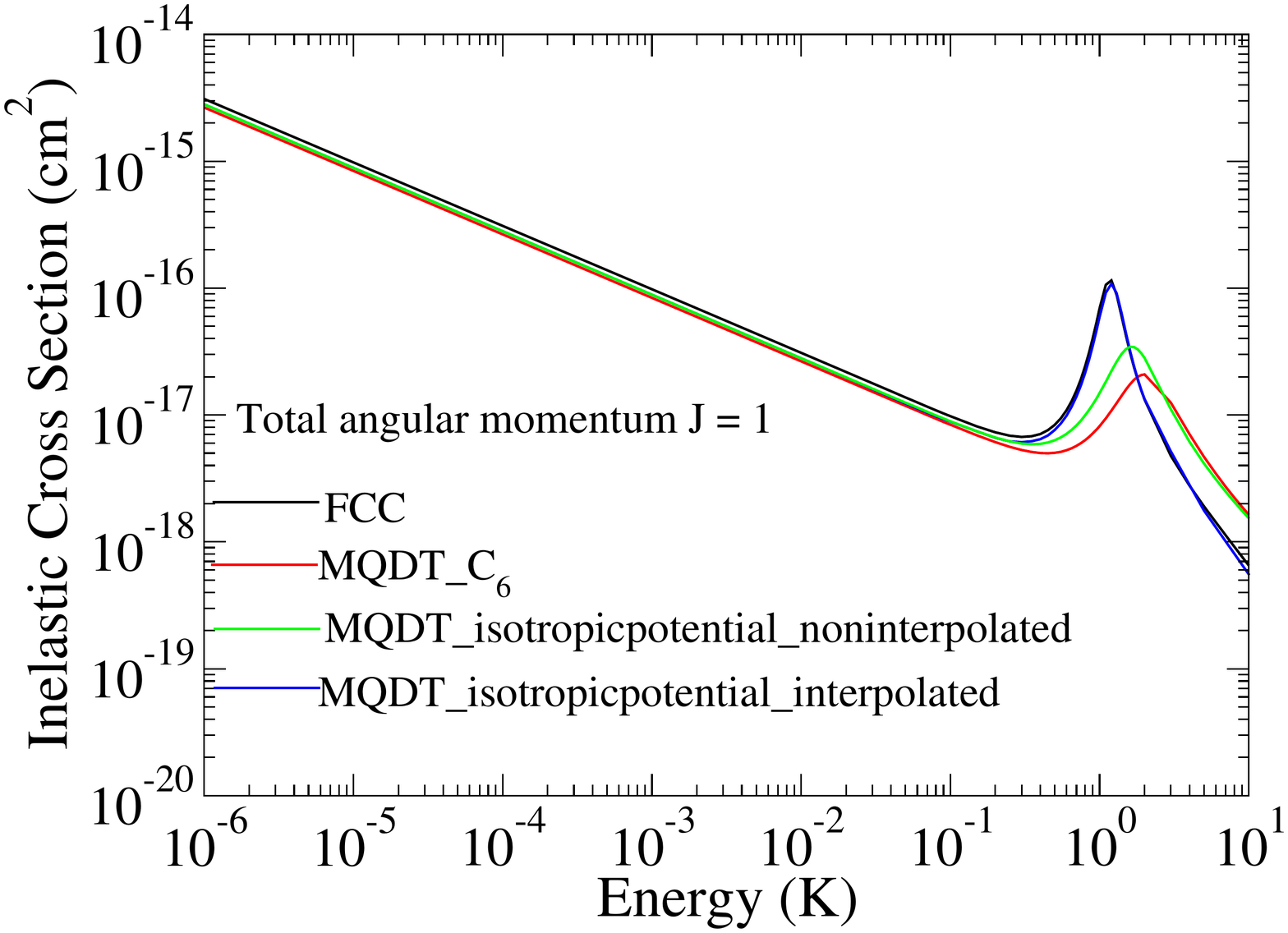}
\end{tabular}
\caption{ Elastic (left) and inelastic (right) cross sections for ortho-para scattering of H$_2$. The black 
curves represent the full CC calculation, 
while the other curves denote different MQDT results as described in the text below.}
\label{OPscatter}
\end{figure}

The resulting cross sections, computed according to the same three degrees of approximation as in Figure \ref{OOscatter},
are presented in Fig.~\ref{OPscatter}, along with that from full CC calculations. The left panel depicts the elastic
cross sections, and the right panel shows the inelastic counterparts. These cross sections were computed for total angular 
momentum $J=1$ 
to account for the 
dominant $s$-wave scattering in the incident channel at ultralow energies.  A matching radius of $R_m = 9.5$ $a_0$ is used
for the MQDT calculations.  

The solid black curve in Fig.~\ref{OPscatter} denotes the full close-coupling results computed on an energy grid of 10$\mu$K.  
The blue curve corresponds to MQDT calculations in which an interpolation scheme similar to that of 
the ortho-ortho case is adopted for the short range ${\bf K}^{\rm sr}$ matrix.
The green curve represents the same calculation, assuming ${\bf K}^{\rm sr}$ (computed at $E_c=1\mu$K) is 
valid at all energies.
Like the ortho-ortho case both the blue (interpolated) and green (non-interpolated) curves almost exactly follow the black curve  in the
Wigner-threshold region, but the green curve begins to deviate above 100 mK as the resonance region is
approached.  

It is abundantly clear from the above discussion that while a  
single ${\bf K}^{\rm sr}$ matrix is not  capable of reproducing  the dynamics over the entire energy range, including the 
resonance region, it is able to accurately describe the dynamics in the Wigner threshold regime.
The accuracy of both the elastic and inelastic cross sections in the three cases considered validates the key idea of MQDT: the energy 
dependence of scattering observables is 
entirely tractable within the simple behavior of the long-range physics.  We also emphasize that the method remains numerically 
tractable all the way down to 1$\mu$ K, about seven orders of magnitude lower in energy than the height of the centrifugal barriers. At 1$\mu$ K (and throughout the Wigner threshold regime) the elastic cross sections from the MQDT and full CC calculations agree
to within 0.2--4\% for all three initial states considered here. The corresponding inelastic cross sections agree to witin 3--10\%.
Finally, we note that the MQDT calculations are also accurate for partial waves $\ell=1,$ and 2 arising from
other values of $J$ (e.g., $J=0,$ 1 for ortho-ortho and $J=0,$ 2 for ortho-para) although they 
do not contribute significantly to the cross sections and hence not shown.

\section{Conclusions}
We have presented a hybrid approach that combines full close-coupling calculations at short-range with
the MQDT formalism at long-range to evaluate cross sections for elastic and quasi-resonant inelastic scattering
in collisions of H$_2$ molecules. It is found that the full CC calculation can be restricted to a relatively
short-range, $\sim$ 9.0 $a.u$, which is just outside the region of the van der Waals potential well. Beyond this
region, the scattering process is described within the MQDT formalism. Further, it is found that
one may use a single short-range $K$-matrix, computed at say $1\mu$K, to evaluate cross sections at energies
all the way into the mK range, leading to significant savings in 
computational time. This works as long as scattering resonances are absent. When resonances are present,
the short-range $K$-matrix becomes sensitive to energy and an interpolation of its elements 
computed  on a relatively
sparse grid in energy may be employed to yield reliable results. 

The choice  of 
H$_2$-H$_2$  for the present work is in part motivated by the possibility of full-dimensional
CC calculations with no approximations (other than basis set truncation). However, due to its shallow
van der Waals potential well, it is probably not the system for which MQDT provides the most accurate values. This is 
because, at the short range matching distance, for collision energies in the Kelvin range, the interaction potential becomes comparable
to the scattering energy and determination of an energy independent short-range K-matrix is no longer
possible. Furthermore, extending the calculations to energies beyond 1 K becomes difficult as 
the effective potential for $\ell>2$ becomes positive for all values of $R$ and 
computation of MQDT reference functions becomes difficult. Nevertheless, the hybrid approach 
seems to be very promising, and the savings in computations will be more dramatic when considering open shell systems 
with spin, hyperfine levels, and magnetic field effects.

While the results demonstrate the relevance of the MQDT approach in ultracold molecule-molecule scattering, there is still much to be developed. 
The treatment of resonances has become mundane in atom-atom scattering, but remains to be adequately developed for cold
molecules \cite{Hud2}. 
Also, the long-range PES for H$_2$-H$_2$ remains reasonably isotropic, so that complete isotropy could be assumed when constructing the reference 
functions.  Potentials with stronger anisotropies may necessitate a different long-range treatment, owing to strong interchannel couplings.  
Finally, for truly reactive systems, such as F+H$_2$, the short-range calculation is more conveniently carried out in hyperspherical, rather than 
Jacobi coordinates, in which case the application of MQDT needs to be modified to accommodate the transition between short- and long-range
coordinate 
systems, as well as between short- and long-range wave functions.  Such calculations are in progress.

\section{Acknowledgements}
This work was supported  in part by NSF grant  PHY-1205838 (N.B.) and
ARO MURI grant No. W911NF-12-1-0476 (N.B. and J.L.B).
Computational
support by National Supercomputing Center for Energy and the Environment at UNLV  is gratefully acknowledged.
JH is grateful to Brian Kendrick for many helpful discussions.

\end{document}